\documentclass[aps,prl,twocolumn,showpacs]{revtex4}

\usepackage{graphicx}
\usepackage{dcolumn}
\usepackage{amsmath}
\begin{document}

\title{Fermi liquid behaviour in an underdoped high $T_{\rm c}$ superconductor}

\author{Suchitra~E.~Sebastian$^1$}
\email{suchitra@phy.cam.ac.uk}
\author{N.~Harrison$^2$}
\email{nharrison@lanl.gov}
\author{M.~M.~Altarawneh$^2$}
\author{Ruixing Liang$^{3,4}$}
\author{D.~A.~Bonn$^{3,4}$}
\author{W.~N.~Hardy$^{3,4}$}
\author{G.~G.~Lonzarich$^1$}

\affiliation{
$^1$Cavendish Laboratory, Cambridge University, JJ Thomson Avenue, Cambridge CB3~OHE, U.K\\
$^2$National High Magnetic Field Laboratory, LANL, Los Alamos, NM 87545\\
$^3$Department of Physics and Astronomy, University of British Columbia, Vancouver V6T 1Z4, Canada\\
$^4$Canadian Institute for Advanced Research, Toronto M5G 1Z8, Canada
}
\date{\today}

\begin{abstract}
We use magnetic quantum oscillations in the underdoped high $T_{\rm c}$ superconductor YBa$_2$Cu$_3$O$_{6+x}$ ($x\approx$~0.56) measured over a broad range of temperatures 100~mK~$<T<$~18~K to extract the form of the distribution function describing the low-lying quasiparticle excitations in high magnetic fields. Despite the proximity of YBa$_2$Cu$_3$O$_{6+x}$ ($x\approx$~0.56) to a Mott insulating state, various broken symmetry ground states and/or states with different quasiparticle statistics, we find that our experimental results can be understood in terms of quasiparticle excitations obeying Fermi-Dirac statistics as in the Landau Fermi liquid theory.
\end{abstract}
\pacs{71.45.Lr, 71.20.Ps, 71.18.+y}
\maketitle

At the heart of our understanding of itinerant electron systems is Landau's Fermi liquid theory in which gapless excitations on a Fermi surface in momentum space are governed by the Fermi-Dirac distribution~\cite{landau1,nozieres1,shoenberg1,ashcroft1}. One of the foremost examples of a breakdown of the Fermi liquid paradigm is in fractional quantum Hall systems~\cite{jain1}. The validity of Landau Fermi liquid theory could also be in question in other interacting systems~\cite{wasserman1} such as those at the border of electron localisation.

The functional form of the Bose-Einstein distribution has been demonstrated in detail by means of a number of experiments including recent optical measurements in cold atomic gases~\cite{anderson2}. Interestingly, however, it is not clear as to what extent of detail the precise form of the Fermi-Dirac distribution function - which is the basis of the theory of electron systems - has been measured~\cite{ashcroft1}. In this paper, we test the validity of Fermi-Dirac statistics governing the elementary excitations in one of the pre-eminent, most strongly correlated examples of a condensed matter system: the underdoped high $T_{\rm c}$ cuprates~\cite{anderson3
}. Quantum oscillations provide a unique means of accessing the statistical distribution governing quasiparticles of an interacting system. While other thermodynamic quantities such as heat capacity are useful in accessing the second moment (i.e. the variance) of the statistical distribution governing the system of interacting particles, we show here the quantum oscillation measurements access not only the second order moment, but also the higher order moments, and hence provide a more complete test of the applicability of the Landau Fermi liquid description.

\begin{figure}
\centering 
\includegraphics*[width=.46\textwidth,angle=0]{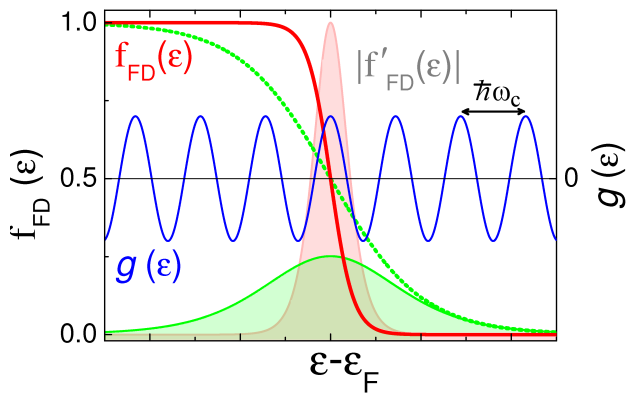}
\caption{Considering the Fermi-Dirac distribution ${\rm f}_{\rm FD}=(1+e^z)^{-1}$ (where $z=[\varepsilon-\varepsilon_{\rm F}]/k_{\rm B}T$), the $T$-dependent step in occupation number (red/green lines) causes the oscillatory density of states $\tilde{g}=g_0e^{i2\pi\varepsilon/\hbar\omega_{\rm c}}$ (assuming that $g_0$ is approximately constant on the scale of the cyclotron energy $\hbar\omega_{\rm c}=\hbar eB/m^\ast$) shown by the blue line to be thermally smeared by the probability distribution $|{\rm f}^\prime_{\rm FD}(z)|=1/2(1+\cosh z)$ (red/green shaded regions). The consequent reduction in amplitude is equivalent to a Fourier transform of $|{\rm f}^\prime_{\rm FD}(z)|$, yielding oscillations $\propto e^{i(2\pi F/B)}$ periodic in $1/B$ (where $F$ is the conventional quantum oscillation frequency~\cite{shoenberg1}) modulated by a $T$-dependent pre-factor $a(T)=a_0\pi\eta/\sinh\pi\eta$ (where $\eta=2\pi k_{\rm B}Tm^\ast/\hbar eB$ and $a_0$ is a constant). The quantum oscillatory magnetisation and resistivity of interest in this paper can be expressed in terms of the above thermally averaged density of states, and hence the same thermal amplitude factor $a(T)$. 
}
\label{doofus}
\end{figure}

We measure magnetic quantum oscillations in YBa$_2$Cu$_3$O$_{6+x}$ ($x\approx$~0.56)~\cite{liang1,doiron1} over an extensive range of temperatures (100~mK~$\leq T\leq$~18~K) to obtain the temperature dependence of the oscillation amplitude $a(T)$. Quantum oscillations originate from changes in the occupancy of partially filled Landau levels at the Fermi energy $\varepsilon_{\rm F}$ as the magnetic induction ${\bf B}$ is varied, making their amplitude particularly sensitive to changes in the thermal distribution function ${\rm f}(z)$ (where $z=[\varepsilon-\varepsilon_{\rm F}]/k_{\rm B}T$)~\cite{shoenberg1}. On increasing the temperature $T$, the broadened probability distribution $|{\rm f}^\prime(z)|=-\partial{\rm f}/\partial z$ smears the phase of the quantum oscillations in energy $\varepsilon$ (see Fig.~\ref{doofus}), causing $a(T)$ to be reduced. The function $a(T)$ measures the Fourier transform of the statistical probability distribution $|{\rm f}^\prime(z)|$. We therefore use the inverse fourier transform of the measured $a(T)$ to access the distribution of interacting quasiparticles in underdoped cuprates. 

\begin{figure}
\centering 
\includegraphics*[width=.35\textwidth,angle=0]{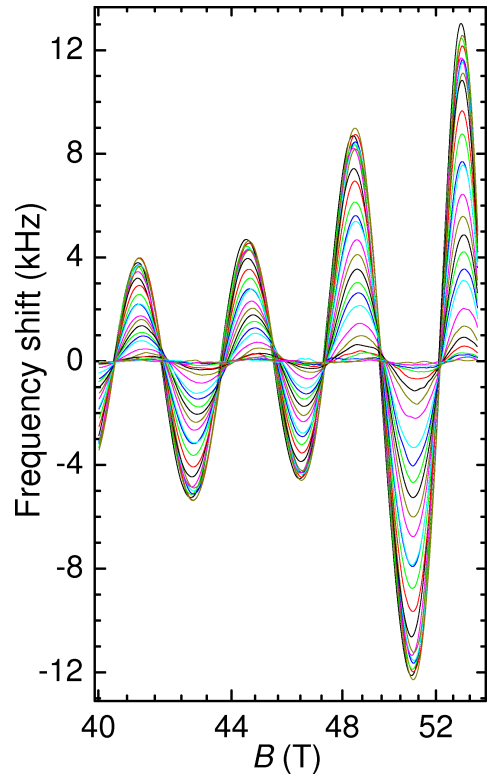}
\caption{Magnetic quantum oscillations measured in YBa$_2$Cu$_3$O$_{6+x}$ with $x\approx$~0.56 (after background polynomial subtraction). This restricted interval in $B=|{\bf B}|$ furnishes a dynamic range of $\sim$~50~dB between $T=$~1 and 18~K. The actual $T$ values are provided in Fig.~\ref{LKplot}.
}
\label{LPdata}
\end{figure}

\begin{figure}
\centering 
\includegraphics*[width=.4\textwidth,angle=-0]{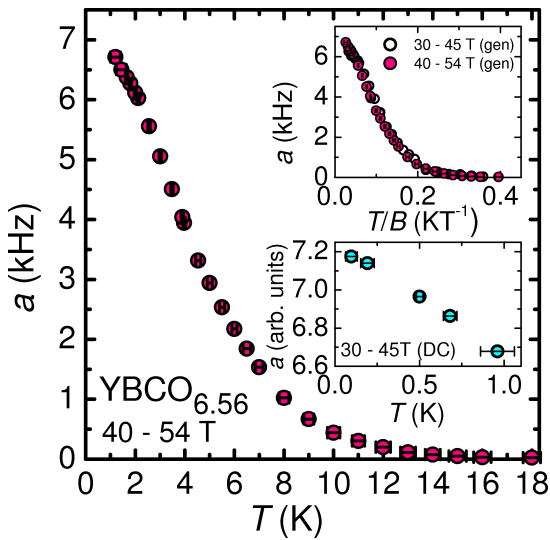}
\caption{Quantum oscillation amplitude $a$ versus $T$ extracted by Fourier analysis of the data in Fig.~\ref{LPdata} for 40~$\leq B\leq$~54~T, corresponding to an average $B=$~45.96~T. 
The upper inset shows the same data (filled circles) plotted versus $T/B$ together with similar points extracted from the field interval 30~$\leq B\leq$~45~T in which the amplitude has been renormalized (open circles), matching the field range of the DC field torque measurements (lower inset). The close correspondence between field ranges implies that the dependence on $T/B$ is independent of the interval in $B$ within the resolution of the experiment, suggesting a magnetic field-independent $m^\ast$. 
}
\label{LKplot}
\end{figure}

An accurate determination of $|{\rm f}^\prime(z)|$ requires $a(T)$ to be measured over a wide range of $T$. Here, we achieve a dynamic range of $\sim$~50~dB (see data in Fig.~\ref{LPdata}) by using the contactless conductivity technique (details described elsewhere~\cite{altarawneh1,sebastian1,sebastian2}) in a motor generator-driven magnet delivering slowly swept magnetic fields to $\sim$~55~T in $^4$He medium down to $\sim$~1~K and torque measurements in a continuous DC magnetic field reaching $\sim$~45~T in a dilution refrigerator. For measurements in the motor generator-driven magnet, the uncertainty in $T$ is minimized by ensuring that the sample is well-coupled to the liquid cryogen (i.e. immersed in liquid $^4$He), thereby minimising heating due to irreversible field effects during the measurements for temperatures below 4.2~K. At these temperatures, the error in $T$ is estimated from a comparison with DC field measurements, and with the resistive crossover field providing an accurate secondary indication of $T$. At temperatures above 4.2~K, the error in $T$ is estimated by repeating field sweeps at different vapour pressures of $^4$He gas. Figure~\ref{LPdata} shows the measured quantum oscillations in resonant frequency shift~\cite{altarawneh1} after polynomial background subtraction at various temperatures; performing a Fourier transformation of the oscillations in $1/B$ using a Hann window yields $a(T)$ of the most prominent ($\alpha$) oscillation shown in Fig.~\ref{LKplot}. Despite the close proximity of two other frequencies~\cite{sebastian3}, the temperature dependence of the $\alpha$ oscillation amplitude is unaffected since the other frequencies are much weaker than the $\alpha$ frequency, and the effective mass of all three frequencies has been measured to be nearly identical~\cite{sebastian3}.

\begin{figure}
\centering 
\includegraphics*[width=.46\textwidth,angle=0]{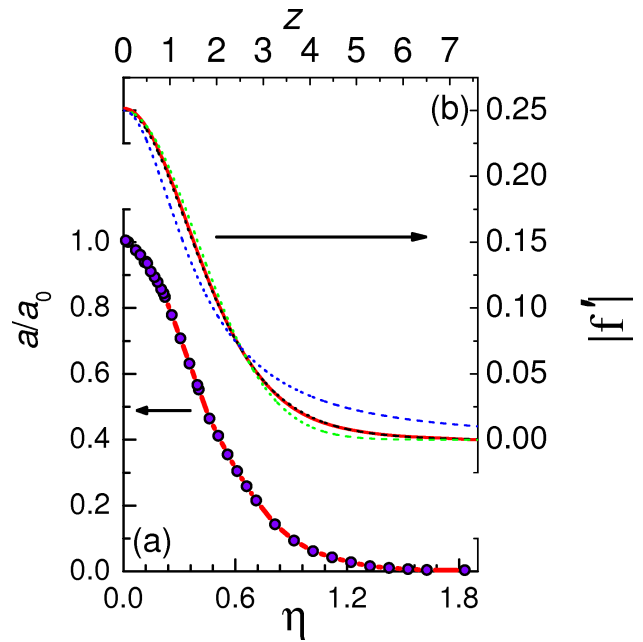}
\caption{{\bf a} Data (circles) from Fig.~\ref{LKplot} plotted versus $\eta=2\pi k_{\rm B}Tm^\ast/eB$, where $m^\ast$=1.676~$m_{\rm e}$. The data from the lower inset to Fig.~\ref{LKplot} is included in this plot after renormalizing its amplitude relative to the contactless conductivity data at $\eta\approx$~0.15. 
The red line corresponds to a linear interpolation. {\bf b} Numerical Fourier transform of $a(\eta)$ in ({\bf a}) (red line) obtained using Eqn.~\ref{FDderivative}, plotted over a restricted range of positive $z=\varepsilon/k_{\rm B}T$, with the Fermi-Dirac (black), Gaussian (green) and Lorentzian (blue) distributions shown for comparison as dotted lines.}
\label{FFT}
\end{figure}

To extend the range in $T$ to lower values, we include low temperature quantum oscillation data measured using magnetic torque in a dilution fridge in DC fields. The temperature error in these dilution fridge measurements could arise either from a nonlinear thermometer magnetoresistance, or a drift in temperature stability during the magnetic field sweep. We calibrate the temperatures in a magnetic field by measuring the magnetoresistance between 11~T and 45~T of the RuO$_2$ thermometer used for these measurements, and collating them with the published values of magnetoresistance up to 8~T~\cite{oxford}. We find the magnetoresistance to behave very close to linearity for all $T\lesssim$~1~K, signalling that the chief error in $T$ arises from thermal drift$-$ estimated from measured changes between rising and falling field.
Although the DC fields extend only up to 45~T (lower inset of Fig.~\ref{LKplot}), the dependence of the amplitude on $T/B$ is found to be independent of the interval in $1/B$ (upper inset of Fig.~\ref{LKplot}), enabling the contactless conductivity and torque data to be combined (upon amplitude renormalization at 1~K) in a plot of the oscillation amplitude versus the dimensionless parameter $\eta=2\pi k_{\rm B}Tm^\ast/\hbar eB$ in Fig.~\ref{FFT}a.

The quasiparticle probability distribution is obtained by performing a Fourier transform of the quantum oscillation amplitude:
\begin{equation}\label{FDderivative}
|{\rm f}^\prime(z)|=\bigg| a_0^{-1}(2\pi)^{-1}\int^{\eta_{\rm lim}}_{-\eta_{\rm lim}}e^{i\eta z}a(T){\rm d}\eta\bigg|.
\end{equation}
Its comparison with the model Fermi-Dirac distribution $|{\rm f}^\prime_{\rm FD}(z)|$ involves two variable parameters: the effective mass $m^\ast$ and the amplitude renormalisation factor $a_0$. These are optimized by fitting to the even partial moments 
\begin{equation}\label{momentevaluation}
\mu_K=\int^{z_{\rm lim}}_{-z_{\rm lim}}z^K|{\rm f}^\prime(z)|{\rm d}z
\end{equation}
to those of the model probability distribution from the lowest order moment upward. While the second moment $\mu_2$ (i.e. variance) can be accessed by other thermodynamic experiments (e.g. heat capacity), 
a key advantage of quantum oscillation measurements is the ability to compare the entire hierarchy of moments with probability distribution models. The chief experimental limitation of quantum oscillation measurements is the breadth of the temperature range over which oscillations are measured. To perform the Fourier transform, evenly spaced points are required in $\pm\eta$. These are obtained by symmetrizing the data in Fig.~\ref{FFT}a with respect to $T=0$ and making a linear interpolation (in the range $-\eta_{\rm lim}<\eta_{\rm m}<\eta_{\rm lim}\approx$~1.8). The data is further padded with zeroes to yield a Fourier transform with points finely spaced in $z$. The combined effects of the finite spacing between data points in Fig.~\ref{LKplot} and experimental uncertainty restrict the reliable range in $z$ to $-z_{\rm lim}<z<z_{\rm lim}\approx$~7.5$-$ hence the comparison of `partial' moments in Table~1 and Eqn.~\ref{momentevaluation}.
%

\begin{table}[h]
\begin{center}
\begin{tabular}{|l|l|l|l|l|l|}
\hline
moment $\mu_K$ & $\mu_2$ & $\mu_4$ & $\mu_6$ & $\mu_8$ & $\mu_{10}$\\
~ & variance & kurtosis & ~ & ~ & ~\\
\hline
${|\rm f}^\prime_{\rm FD}|$~($\pm\infty$) & 3.2899 & 45.458 & 1419 & 8.03$\times$10$^4$ & 7.25$\times$10$^6$\\
\hline
${|\rm f}^\prime_{\rm FD}|$~($\pm z_{\rm lim}$) & 3.2089 & 39.123 & 875 & 2.70$\times$10$^4$ & 9.94$\times$10$^5$\\
\hline
${|\rm f}^\prime_{\rm exp}|$~($\pm z_{\rm lim}$) & 3.2085 & 39.128 & 873 & 2.65$\times$10$^4$ & 9.49$\times$10$^5$\\
\hline
\end{tabular}
\caption{\label{momentstable} Evaluation of the lowest even moments. The first row contains the lowest even moments of the Fermi-Dirac probability (corresponding to $z_{\rm lim}=\infty$). The second row contains the corresponding partial moments of the Fermi-Dirac probability distribution using $z_{\rm lim}=$~7.5 identical to the experimental data limits for purposes of comparison, and the third row contains the partial moments evaluated for the probability distribution extracted from the experimental data also using $z_{\rm lim}=$~7.5. Very good agreement is seen.}
\end{center}
\vspace{-0.6cm}
\end{table}

\begin{figure}
\centering 
\includegraphics*[width=.46\textwidth,angle=0]{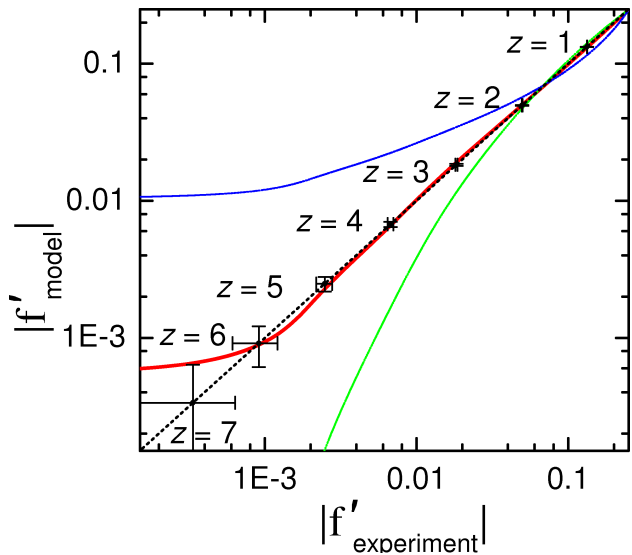}
\caption{{\bf a} Comparison of the fit probability distributions shown in Fig.~\ref{FFT} with that obtained from experiment. The Fermi-Dirac (red line), Gaussian (green dotted line) and Lorentzian (blue dotted line) probability distributions are shown on the y-axis as a function of the experimentally obstained probability distribution $|{\rm f}^\prime(z)|$ on the x-axis. The circles locate integer values of $z$ (the black dotted line is a guide to the eye), with the error bars indicating the extent to which experimental uncertainties (chiefly associated with the sample temperature) can cause $|{\rm f}^\prime(z)|$ to depart from $|{\rm f}^\prime_{\rm FD}(z)|$. }
\label{prob}
\end{figure}

On adopting this procedure of multiple moment comparison for $-z_{\rm lim} < z < z_{\rm lim}$, we find optimised values of $a_0$ and $m^\ast=$~1.676~$\pm$~0.001 for which the second (variance) and fourth (kurtosis) moments of the experimentally obtained probability distribution concides to 4 and 3 significant digits with the Fermi-Dirac probability distribution, respectively. Good agreement continues to be found with the higher order moments, which we list in Table I up to $\mu_{10}$. In Fig.~\ref{prob} we compare the Fermi-Dirac distribution and the illustrative Gaussian and Lorentzian probability distributions against that obtained experimentally. This exercise demonstrates the importance of a wide temperature range in testing the goodness of fit of a probability distribution model$-$ we see that the deviation from the Gaussian and Lorentzian models only appears above $z=2$. The deviation from an ideal Fermi-Dirac distribution occurs only above $z=6$, but falls within the experimental uncertainty.

Our study therefore indicates a high degree of correspondence between the elementary excitations in underdoped YBa$_2$Cu$_3$O$_{6+x}$ ($x\approx$~0.56) and those of fermionic quasiparticles governed by a Fermi-Dirac distribution. Hence, long-lived, robust fermionic quasiparticles exist over a broad range of temperatures (100~mK~$<T<$~18~K). 
At first glance one may find this result surprising, especially in a system such as underdoped YBa$_2$Cu$_3$O$_{6+x}$ which is close to the Mott insulating state. We relook at whether we indeed have reason to find this result unexpected, and investigate possible reasons why Landau Fermi liquid behaviour prevails. We consider alternative scenarios that may be applicable to the ground state of underdoped YBa$_2$Cu$_3$O$_{6+x}$, and investigate whether they would be consistent with our experimental observation of Fermi-Dirac statistics. In addition to non Fermi-Dirac statistics, the temperature dependence of the oscillation amplitude could deviate from that shown in Fig.~\ref{doofus} due to a non-constant value of $g_0$. An energy gap in the electronic structure near the Fermi energy of magnitude comparable to $\hbar\omega_{\rm c}$ could yield a non-constant value of $g_0$, such as in the case of symmetry-breaking groundstates (e.g. density wave ordered states with Fermi surface reconstruction due to translational symmetry breaking and the superconducting vortex liquid). Any additional temperature dependence of the amplitude arising from these effects, however, appears not to be large enough to lead to a detectable deviation in our experiment from the temperature-dependent amplitude expected from Fermi-Dirac statistics.

Various other models would not be expected to lead to Fermi-Dirac statistics - a notable example being fractional quantum Hall effect in very high magnetic fields. In the case of underdoped YBa$_2$Cu$_3$O$_{6+x}$, models yielding oscillations unrelated to Landau quantisation (e.g.~Ref.~\cite{alex}) would not be expected to result in a temperature dependent oscillation amplitude that agrees with Fermi-Dirac statistics, and are therefore inconsistent with our experimental findings. To consider the applicability of various other non Fermi liquid models proposed for YBa$_2$Cu$_3$O$_{6+x}$~(e.g.~Ref.~\cite{anderson3}), theoretical predictions are required for the energy scale at which a deviation from Fermi-Dirac statistics is expected. Such ground states could potentially go undetected in the current study if the deviation from Fermi-Dirac statistics takes place over an energy scale lower than that $\sim$~100~mK of the lowest accessed temperature. Similar considerations may apply to the superconducting vortex liquid groundstate in which regime the current experiments are performed. It will be of interest to determine conditions under which a breakdown of Fermi-Dirac statistics could occur for this groundstate.

This work is supported by the US Department of Energy, the National Science Foundation, the State of Florida, the Royal Society, and Trinity College (University of Cambridge).

\end{document}